# A Novel Estimation Method for Temperature of Magnetic Nanoparticles Dominated by Brownian Relaxation Based on Magnetic Particle Spectroscopy

Zhongzhou Du, Gaoli Zhao, Zhanpeng Hua, Na Ye, Yi Sun, Wenjie Wu, Haochen Zhang, Longtu Yu, Shijie Han, Haozhe Wang, Wenzhong Liu, *Member, IEEE*, and Takashi Yoshida, *Member, IEEE*

***Abstract*—This paper presents a novel method for estimating the temperature of magnetic nanoparticles (MNPs) based on AC magnetization harmonics of MNPs dominated by Brownian relaxation. The difference in the AC magnetization response and magnetization harmonic between the Fokker-Planck equation and the Langevin function was analyzed, and we studied the relationship between the magnetization harmonic and the key factors, such as Brownian relaxation time, temperature, magnetic field strength, core size and hydrodynamic size of MNPs, excitation frequency, and so on. We proposed a compensation function for AC magnetization harmonic with consideration of the key factors and the difference between the Fokker-Planck equation and the Langevin function. Then a temperature estimation model based on the compensation function and the Langevin function was established. By employing the least squares algorithm, the temperature was successfully calculated. The experimental results show that the temperature error is less than 0.035 K in the temperature range from 310 K to 320 K. The temperature estimation model is expected to improve the performance of the magnetic nanoparticle thermometer and be applied to magnetic nanoparticle-mediated hyperthermia.**

***Index Terms*—magnetic nanoparticle, Fokker-Planck equation, Brownian relaxation, Langevin function.**

## I. Introduction

MAGNETIC nanoparticle-mediated hyperthermia (MNPH) [1], [2], [3] is a new anticancer therapy that heats local body parts to kill cancer cells using the difference in heat resistance between tumor tissue cells and normal cells. Magnetic nanoparticles (MNPs) induce damage or necrosis of cancerous cells by elevating their temperature above 315 K–319 K (42 °C–46 °C) without significantly harming the surrounding healthy tissue. The temperature of tumor-affected tissue is necessary to monitor for tissue necrosis and leave the adjacent healthy tissue undamaged [4], [5], [6], [7]. Temperature is an important parameter that can reflect the state of internal physiological conditions. Determining the temperature of living cells or tissues in vivo can enable the development of diagnostic and therapeutic techniques for some cancers [8], [9], [10].

Magnetic nanoparticle thermometer (MNPT) is a new tool that non-invasively measures temperature using the temperature dependency of the nonlinear magnetization of MNPs [11], [12], [13], [14], [15], [16]. Liu *et al.* [17] used the temperature sensitivities of the amplitudes of the first and third harmonics of MNP magnetization by an AC magnetic field to measure the temperature of MNPs. Pi *et al.* [18] studied both odd and even harmonics of MNPs based on the Langevin function under low frequency (117 Hz) AC and DC magnetic fields, and proposed mathematical models of weighted amplitude summation of both odd and even harmonics of MNPs. The temperature of MNPs was solved using the first two terms of Taylor's expansion of the Langevin function under low-frequency (25 Hz) and weak triangular-wave applied AC magnetic field by Zhong *et al.* [19]. In a previous study, the temperature measurement and feedback control system based on an improved MNPT achieves non-invasive temperature sensing using MNP temperature measurement [16]. The frequency of the excited field heating the MNPs is up to 100 kHz, while the temperature model just applies a low-frequency magnetic field (less than 1 kHz). The temperature of MNPs cannot be measured when heated in the system, which can be solved by improving the excitation frequency range of the temperature model gradually. During the improvement process, we are faced with the problem of Brownian relaxation. The empirical expression for harmonics of AC magnetization of MNPs dominated by Brownian relaxation for MNPT was studied in our previous research [20]. The empirical expression is suitable when the effect of Brownian relaxation is negligible or not significant. The influence of Brownian relaxation is quite neglected or not fully considered in previous studies. Brownian relaxation is always present in MNPs exposed to AC excitation fields, and the Langevin function cannot accurately describe the

This work was supported by National Key Research and Development Program of China under Grant 2022YFE0107500, Natural Science Foundation of Henan under Grant 232300420149, and China-Belt and Road Joint Laboratory On Measurement and Control Technology under Grant MCT202304. *(Corresponding author: Na Ye.)*

Zhongzhou Du, Gaoli Zhao, Zhanpeng Hua, Na Ye, and Wenjie Wu are with the School of Computer Science and Technology, Zhengzhou University of Light Industry, Zhengzhou 450001, China (e-mail: duzhongzhou1983@gmail.Com; gaolizhao0926@gmail.com; zhanpenghua616@gmail.com; yena@zzuli.edu.cn; 542107090220@email.zzuli.edu.cn).

Yi Sun, Haochen Zhang, Longtu Yu, Haozhe Wang, and Takashi Yoshida are with the Department of Electrical and Electronic Engineering, Kyushu University, Fukuoka 819-0395, Japan (e-mail: sunyi3064@gmail.com; zhang.haochen.048@s.kyushu-u.ac.jp; yulongtug@gmail.com; wang.haozhe.726@s.kyushu-u.ac.jp; t_yoshi@ees.kyushu-u.ac.jp).

Shijie Han and Wenzhong Liu are with the School of Artificial Intelligence and Automation, Huazhong University of Science and Technology, Wuhan 430074, China (e-mail: hanshijie@hust.edu.cn; lwz7410@hust.edu.cn).

AC magnetization dynamics of MNPs dominated by Brownian relaxation. Brownian relaxation has a crucial impact on the measurement accuracy and time resolution of MNPT, and limits the application of MNPT.

In this study, we aimed to find a new estimation method for the temperature of MNPs dominated by Brownian relaxation. We analyzed the difference in the AC magnetization response and magnetization harmonic between the Fokker-Planck equation and the Langevin function, and studied the relationship between the harmonic amplitude and the key factors, the key factors include Brownian relaxation time, temperature, magnetic field strength, core size and hydrodynamic size of MNPs, excitation frequency, and so on. Based on the difference between the Fokker-Planck equation and the Langevin function, and the key factors were also taken into account, the compensation function was established. Then a temperature estimation model based on the compensation function and the Langevin function was established, and the temperature was solved using the least squares algorithm.

## II. Model And Method

### A. The Langevin Function

MNPs were exposed to an AC magnetic field with a sufficiently low frequency, and the magnetization can be described by the Langevin function ignoring Brownian relaxation:

$$M_L(t,\omega,H,T)=\phi M_S\left(\coth\left(\frac{M_S V_0 H}{K_B T}\right)-\frac{K_B T}{M_S V_0 H}\right) \quad (1)$$

where $\phi$ is the concentration of the MNP-based sample, $H = \mu_0 H_0\cos(\omega t)$ is the excitation field, $\mu_0 H_0$ is the magnetic field strength, $\mu_0$ is the magnetic permeability of vacuum, $\omega = 2\pi f$ is the angular frequency, $f$ is the excitation frequency, $M_s$ is the saturation magnetization, $K_B$ is the Boltzmann constant, $V_0 = \pi d_c^3/6$ is the volume of a particle and $T$ is the absolute temperature. Taylor series expansion of (1) allows $M_L(t, \omega, H, T)$ to be expressed as:

$$M_L(t,\omega,H,T)=\sum_{j=1}^{\infty}A_{2j-1}(\omega,H,T)\sin((2j-1)\omega t) \quad (2)$$

where $A_{2j-1}$ is the (2j-1)-th harmonic amplitude of magnetization.

### B. Fokker-Planck Equation for Brownian Relaxation

The dynamics of magnetization of MNPs dominated by Brownian relaxation can be accurately described by the Fokker-Planck equation:

$$2\tau_{B,0}\frac{\partial W(\theta,t)}{\partial t}=\frac{1}{\sin\theta}\frac{\partial}{\partial\theta}\left\{\sin\theta\left[\xi W(\theta,t)\sin\theta+\frac{\partial W(\theta,t)}{\partial t}\right]\right\} \quad (3)$$

where $\theta$ is the angle of the magnetic moment $m$ for the AC applied field $H$, $W(\theta, t)$ is the distribution function of $\theta$, and $\xi = mH/K_BT$. $\tau_{B,0} = \pi\eta d_h^3/2K_BT$ is the Brownian relaxation, $\eta$ is the viscosity coefficient, and $d_h$ is the hydrodynamic size of MNPs. The magnetization response of MNPs $M_{FP}(t, \omega, H, \tau_{B,0}, T)$ in the direction of $H$ can be described as:

$$M_{FP}(t,\omega,H,\tau_{B,0},T)=\phi M_s\int_0^{\pi}W\sin\theta\cos\theta d\theta \quad (4)$$

Fourier series expansion of (4) allows $M_{FP}(t, \omega, H, \tau_{B,0}, T)$ to be expressed as:

$$M_{FP}(t,\omega,H,\tau_{B,0},T)=\sum_{j=1}^{\infty}\left[B_{2j-1}(\omega,H,T)\sin((2j-1)\omega t+\varphi_{2j-1})\right] \quad (5)$$

where $B_{2j-1}$ and $\varphi_{2j-1}$ are the (2j-1)-th harmonic amplitude and phase of magnetization, respectively.

### C. The Compensation Function for Magnetization Harmonic of MNPs Dominated by Brownian Relaxation

It is easy to solve for the analytical expression of harmonic amplitude and temperature based on the Langevin function without consideration of Brownian relaxation. The AC magnetization dynamics of MNPs dominated by Brownian relaxation can be described accurately utilizing the Fokker-Planck equation, but the AC magnetization response of MNPs calculated via the Fokker-Planck equation is extremely complicated, and the analytical harmonic expression for MNPT is hard to solve.

We try to construct a new function to describe the magnetization harmonic of MNPs dominated by Brownian relaxation, which was established using the Langevin function and a compensation function. The compensation function was established by the difference between the Fokker-Planck equation and the Langevin function, and also took into account the key factors, including Brownian relaxation time, temperature, magnetic field strength, core size and hydrodynamic size of MNPs, excitation frequency, and so on. The relationship between the Fokker-Planck equation and the Langevin function can be expressed as:

$$M_{FP}(t,\omega,H,\tau_{B,0},T)=\sum_{j=1}^{\infty}\left[G_{2j-1}(\omega,H_0,\tau_{B,0})A_{2j-1}(\omega,H,T)\sin\left((2j-1)\omega t+\varphi_{2j-1}\right)\right] \quad (6)$$

$$G_{2j-1}(\omega,H_0,\tau_{B,0})=1\Bigg/\sqrt{1+((2j-1)\omega\tau_{B,0})^2\frac{1}{1+a_{2j-1}\gamma^2}} \quad (7)$$

where $G_{2j-1}(\omega, \mu_0H_0, \tau_{B,0})$ is the compensation function, $\gamma = \mu_0H_0M_s\pi d_c^3/6K_BT$, $a_{2j-1}$ is related to the MNP-based sample and the order of magnetization harmonic. The relationship between $B_{2j-1}$ and $A_{2j-1}$ could be obtained from (5) and (6):

$$B_{2j-1}(\omega,H,T)=G_{2j-1}(\omega,H_0,\tau_{B,0})A_{2j-1}(\omega,H,T) \quad (8)$$

To eliminate the concentration of the MNP-based sample, the amplitudes of the first and third harmonics were used to establish the model:

$$\begin{cases}B_1=G_1\left(\frac{xy}{3}-\frac{xy^3}{60}+\frac{xy^5}{756}-\frac{xy^7}{8640}+\frac{xy^9}{95040}\cdots\right)\\ B_3=G_3\left(\frac{xy^3}{180}-\frac{xy^5}{1512}+\frac{xy^7}{14400}-\frac{xy^9}{142560}\cdots\right)\end{cases} \quad (9)$$

where $x = \phi M_s$, $y = \mu_0H_0M_sV_0/K_BT$, and $T$ is the temperature to be measured. $B_1$ and $B_3$ are the amplitudes of the first and third harmonics of MNPs dominated by Brownian relaxation, respectively. Temperature of MNPs is obtained from (9) using the least squares algorithm.

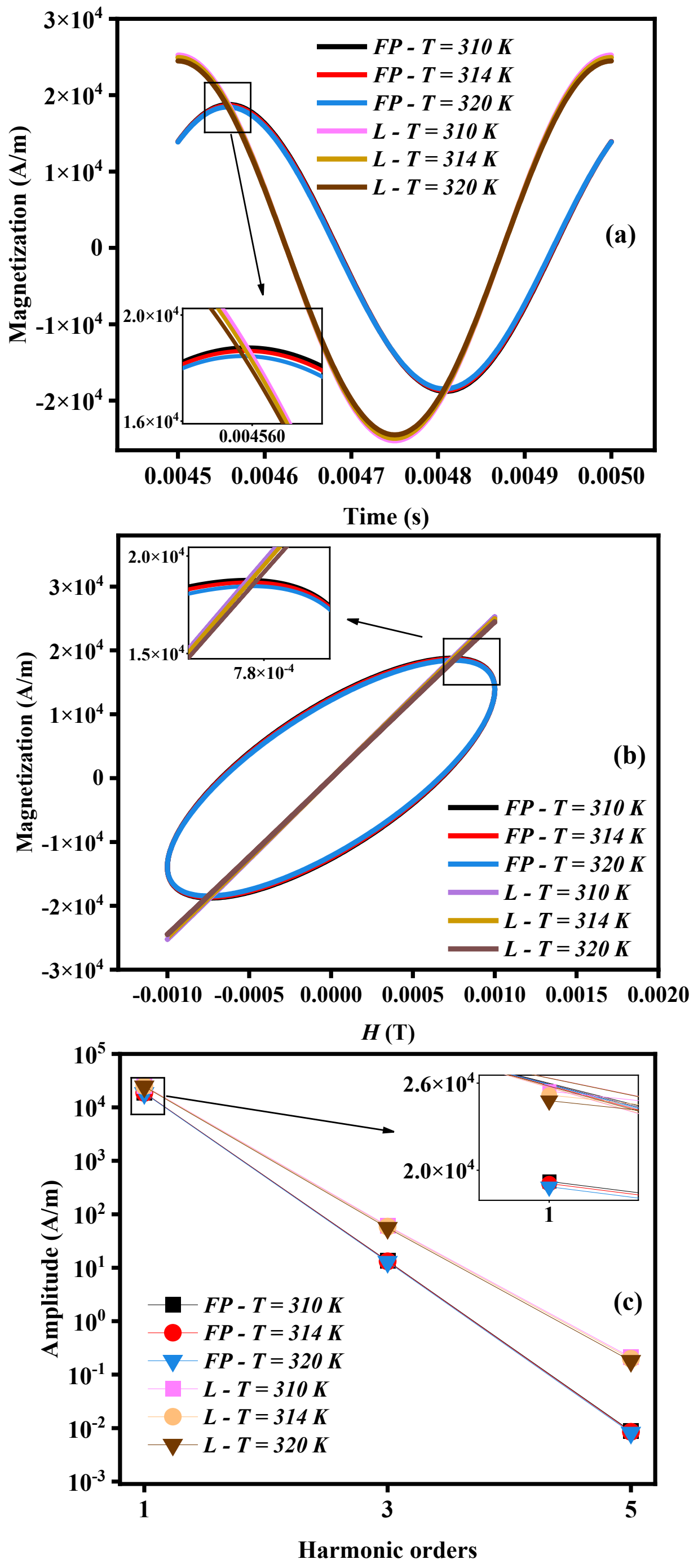

**Fig. 1.** The difference in the magnetization response and harmonic amplitude between the Fokker-Planck equation and the Langevin function under different temperatures ($T$ = 310 K, 314 K, and 320 K). (a) magnetization response of MNPs, (b) M-H curves, and (c) the amplitudes of the first, third, and fifth harmonics. The parameters for the simulation are $\mu_0 H_0$ = 0.001 T, $d_c$ = 25 nm, $f$ = 2 kHz, $d_h$=60 nm, $M_s$ = 200 kA/m, $\eta$ = 0.914 $\times 10^{-3}$ Pas.

## III. Simulation

We performed simulations utilizing MATLAB software to determine the difference in the magnetization response and harmonic amplitude between the Fokker-Planck equation and the Langevin function.

Simulation was carried out under different temperatures to determine the difference in the stable magnetization responses between the Fokker-Planck equation and the Langevin function, and study the relationship among the temperature, the magnetization response, and harmonic amplitude. In the simulation, the excitation frequency and magnetic field strength were 2 kHz and 0.001 T respectively. The MNPs were assumed to have a core diameter of 25 nm without the core size distribution, and the hydrodynamic size was 60 nm. The saturation magnetization was set to 200 kA/m, and the temperature was set to 310 K, 314 K, and 320 K.

As shown in Fig. 1(a) and Fig. 1(b), the magnetization responses based on the Fokker-Planck equation were calculated from (4) under temperatures of 310 K (black lines), 314 K (red lines), and (blue lines) respectively, and the magnetization responses based on the Langevin function were calculated from (1) under temperatures of 310 K (pink lines), 314 K (yellow lines), and (brown lines) respectively. As depicted by the curves in Fig. 1(a), the difference between the magnetization responses calculated from (1) and (4) is in the amplitude and time delay. In the inset of Fig. 1(a), the higher the temperature, the smaller the magnetization response of MNPs. Then we compared the M-H curves in Fig. 1(b). The M-H curves based on the Langevin function have no hysteresis loop because the Brownian relaxation is completely ignored. Due to the influence of Brownian relaxation, the magnetization response based on the Fokker-Planck equation has a hysteresis loop in the M-H curve. Magnetization response decreases with increasing temperature in the inset of Fig. 1(b). The harmonic amplitudes were calculated from the magnetization responses based on the Fokker-Planck equation using the digital phase sensitive detection (DPSD) algorithm under temperatures of 310 K (black squares), 314 K (red circles), and 320 K (blue triangles) in Fig. 1(c). Pink squares, yellow circles, and brown triangles represent the harmonic amplitudes calculated from the magnetization responses based on the Langevin function under temperatures of 310 K, 314 K, and 320 K respectively. In the inset of Fig. 1(c), the higher the temperature, the smaller the amplitudes of the first, third, and fifth harmonics. The amplitudes of the first, third, and fifth harmonics based on the Fokker-Planck equation are smaller than those of the Langevin function due to the influence of Brownian relaxation.

There is a clear difference between the Fokker-Planck equation and the Langevin function due to the influence of Brownian relaxation time in Fig. 1. Next, we performed simulations with different Brownian relaxation times to determine the difference between the stable magnetization responses calculated from (1) and (4), and studied the relationship among the Brownian relaxation time, the magnetization response, and harmonic amplitude. In the simulation, the excitation frequency was 2 kHz and the magnetic field strength was 0.001 T. The MNPs were assumed to have a core diameter of 25 nm without the core size distribution, the Brownian relaxation times were set to 7.2475e-5 s, 1.1508e-4 s, and 1.7179e-4 s. The saturation magnetization was set to 200 kA/m, and the temperature was set to 297 K. In Fig. 2(a) and 2(b), the magnetization responses were calculated from (4) under Brownian relaxation times of 7.2475e-5 s (black lines), 1.1508e-4 s (red lines), and 1.7179e-4 s (blue lines) respectively. The pink lines represent the magnetization responses calculated from (1).

As shown in Fig. 2(a), the longer the Brownian relaxation time, the greater the difference in the magnetization response between the Fokker-Planck equation and the Langevin function. Fig. 2(b) displays the M-H curves of MNPs. The magnetization response based on the Fokker-Planck equation has a hysteresis loop in the M-H curve, which indicates that the magnetization response of MNPs with Brownian relaxation delays the excitation field. Then the harmonic amplitudes based on the Fokker-Planck equation were calculated using the DPSD algorithm under Brownian relaxation times of 7.2475e-5 s (black squares), 1.1508e-4 s (red circles), and 1.7179e-4 s (blue triangles) respectively in Fig. 2(c). Pink triangles are the harmonic amplitudes calculated from the magnetization responses based on the Langevin function. The amplitudes of the first, third, and fifth harmonic diminishes as Brownian relaxation time increases. This effect is particularly pronounced for the higher-order harmonics, where the amplitude demonstrates a more noticeable decline with escalating Brownian relaxation time as the harmonic order increases.

The simulations were carried out to further study the process by which the factors affect Brownian relaxation time and harmonic amplitude, such as magnetic field strength, core size of MNPs, hydrodynamic size of MNPs, and excitation frequency. The harmonic amplitude was calculated using the DPSD algorithm.

The simulation was conducted under different magnetic field strengths to study the relationship among magnetic field strength, Brownian relaxation time, and harmonic amplitude. The relationship among Brownian relaxation time, the magnetic field strength, and the core size of MNPs can be expressed as $\tau_B(\mu_0 H_0, d_c) = \tau_{B,0}/(1+0.07\times\zeta^2)^{0.5}$, and $\zeta = \mu_0 H_0 M_s \pi d_c^3/6K_B T$ [21], [22]. For the calculation of the amplitudes of the first and third harmonics, the AC excitation field had a frequency of 2 kHz, and the magnetic field strength was set from 0.001 T to 0.009 T with a step of 0.002 T. The MNPs were assumed that there was no core size distribution, and the core diameter of MNPs was 25 nm. The hydrodynamic size was set to 60 nm, and the saturation magnetization was 200 kA/m. Fig. 3(a) shows the harmonic amplitudes under different magnetic field strengths. Black and red squares represent the amplitudes of the first and third harmonics obtained from the Fokker-Planck equation, respectively. Black and red circles represent the amplitudes of the first and third harmonics calculated from the magnetization responses based on the Langevin function, respectively. The inset of Fig. 3(a) shows the effect of magnetic field strength on Brownian relaxation time, and Brownian relaxation time was calculated from $\tau_B(\mu_0 H_0, d_c)$. As shown in Fig. 3(a), Brownian relaxation time decreases with increasing magnetic field strength. The greater the magnetic field strength, the shorter the Brownian relaxation time, and the weaker the effect of Brownian relaxation on the amplitudes of the first and third harmonics.

The simulation was performed with different core sizes of MNPs to analyze the relationship among the core size of MNPs, Brownian relaxation time, and harmonic amplitude. In the simulation, the excitation frequency was 2 kHz and the magnetic field strength was 0.001 T. The core sizes of MNPs were set from 18 nm to 30 nm with a step of 3 nm. The hydrodynamic size of MNPs was 60 nm and the saturation magnetization was 200 kA/m. Fig. 3(b) shows the harmonic amplitudes under different core sizes of MNPs. The inset of Fig. 3(b) shows the relationship between the core size of MNPs and Brownian relaxation time, and Brownian relaxation time was calculated from $\tau_B(\mu_0 H_0,\ d_c)$. Brownian relaxation time decreases with increasing core size of MNPs. In Fig. 3(b), black and red squares represent the amplitudes of the first and third harmonics obtained from the Fokker-Planck equation, respectively. Black and red circles represent the amplitudes of the first and third harmonics obtained from the Langevin function, respectively. The greater the core size of MNPs, the shorter the Brownian relaxation time, and the weaker the effect of Brownian relaxation on the amplitudes of the first and third harmonics. The difference in the amplitudes of the first and third harmonics between the Fokker-Planck equation and the Langevin function decreases with increasing core size of MNPs.

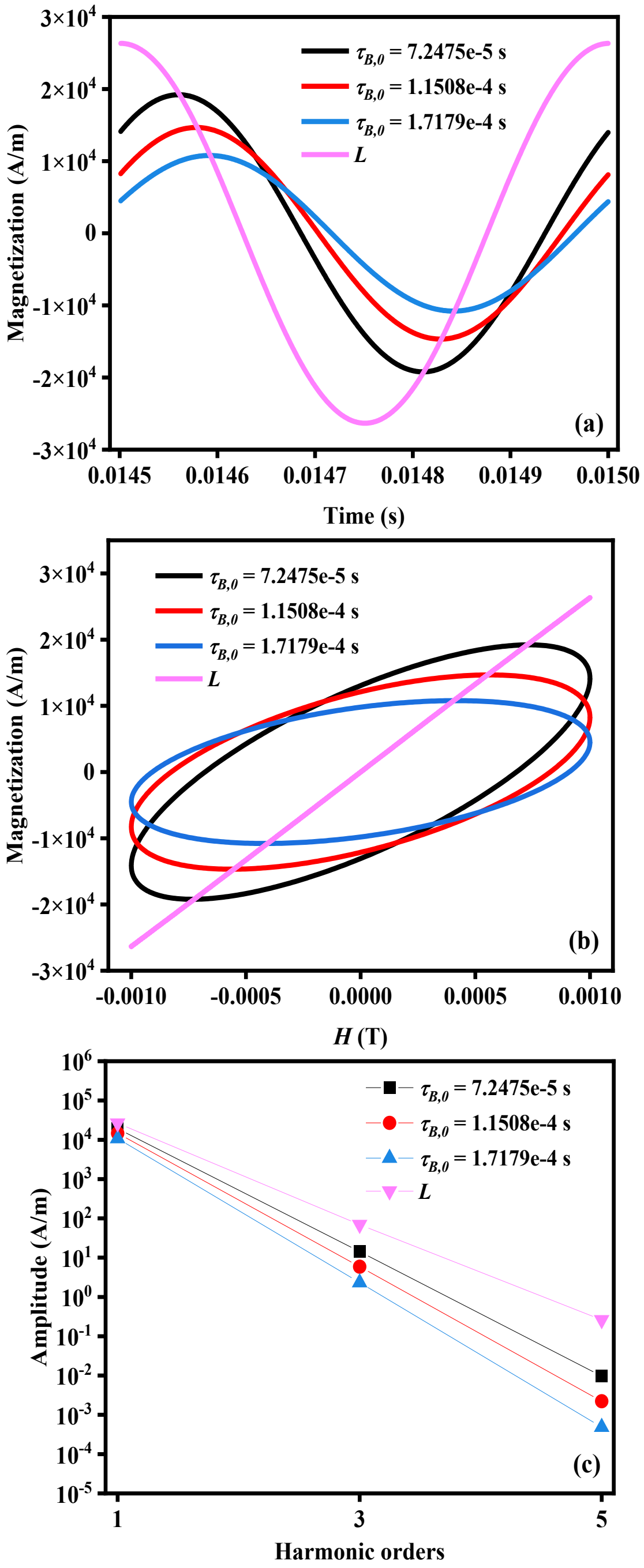

**Fig. 2.** The difference in the magnetization response and harmonic amplitude between the Fokker-Planck equation and the Langevin function under different Brownian relaxation times ($\tau_{B,0}$ = 7.2475e-5 s, 1.1508e-4 s, and 1.7179e-4 s). (a) magnetization response of MNPs, (b) the M-H curve, and (c) the amplitudes of the first, third, and fifth harmonics. The parameters for the simulation are $\mu_0 H_0$ = 0.001 T, $f$ = 2 kHz, $d_c$ = 25 nm, $T$ = 297 K, $M_s$ = 200 kA/m, $\eta$ = 0.914×$10^{-3}$ Pas.

To study the relationship among the hydrodynamic size of MNPs, Brownian relaxation time, and harmonic amplitude, the simulation was carried out under different hydrodynamic sizes of MNPs. In the simulation, the frequency and amplitude of the AC excitation field were 2 kHz and 0.001 T, respectively. The MNPs were assumed that there was no core size distribution, and the core diameter of MNPs was 25 nm. The hydrodynamic sizes of MNPs were set from 60 nm to 100 nm with a step of 10 nm, and the saturation magnetization was 200 kA/m. The inset of Fig. 3(c) shows the relationship between the hydrodynamic size of MNPs and Brownian relaxation time, and Brownian relaxation time was calculated from $\tau_{B,0}$. Brownian relaxation time increases with increasing hydrodynamic size of MNPs. Fig. 3(c) shows the amplitudes of the first and third harmonics under different hydrodynamic sizes of MNPs. Black and red squares represent the amplitudes of the first and third harmonics obtained from the Fokker-Planck equation respectively. The black and red circles represent the amplitudes of the first and third harmonics obtained from the Langevin function, respectively. The amplitudes of the first and third harmonics based on the Langevin function are independent of the hydrodynamic size of MNPs. The difference in the amplitudes of the first and third harmonics between the Fokker-Planck equation and the Langevin function grows with increasing hydrodynamic size of MNPs.

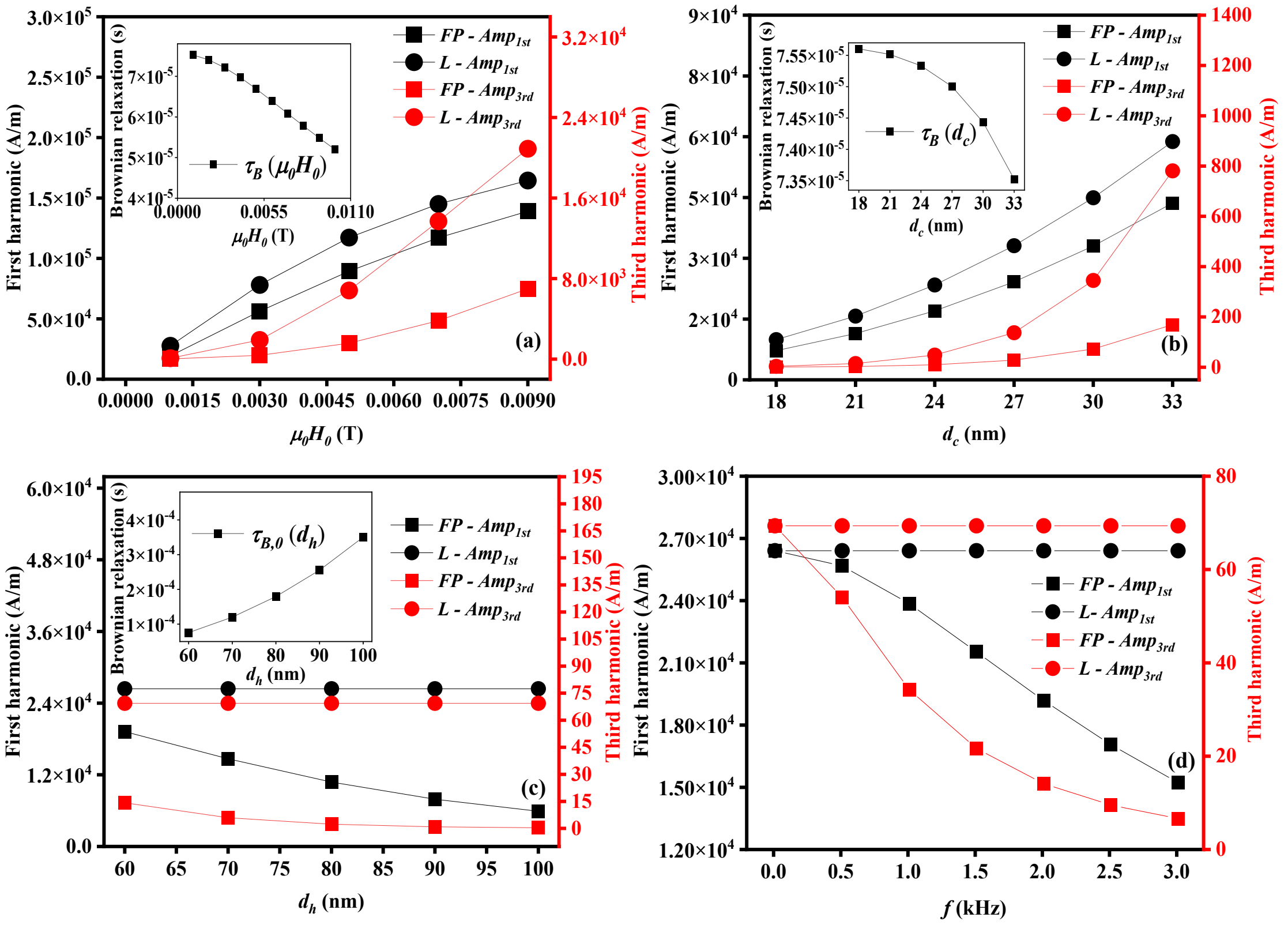

**Fig. 3.** The harmonic amplitudes respectively obtained from the Fokker-Planck equation and the Langevin function under different magnetic field strengths, core sizes of MNPs, hydrodynamic sizes of MNPs, and excitation frequencies. The insets show the relationship between the factors and Brownian relaxation time. (a) the magnetic field strengths were set from 0.001 T to 0.009 T with a step of 0.002 T, (b) the core sizes of MNPs were set from 18 nm to 30 nm with a step of 3 nm, (c) the hydrodynamic sizes of MNPs were set from 60 nm to 100 nm with a step of 10 nm, and (d) the excitation frequencies were set from 0.01 kHz to 3.01 kHz with a step of 0.5 kHz.

We performed the simulation under different excitation frequencies to analyze the relationship among excitation frequency, Brownian relaxation time, and harmonic amplitude. The following parameters were used in the simulation, the magnetic field strength was set to 0.001 T, the excitation frequency was set from 0.01 kHz to 3.01 kHz with a step of 0.5 kHz, the MNPs were assumed there was no core size distribution, and had a core diameter of 25 nm. The hydrodynamic size of MNPs was set to 60 nm, and the saturation magnetization was 200 kA/m. The harmonic amplitudes under different excitation frequencies are shown in Fig. 3(d). Black and red squares represent the amplitudes of the first and third harmonics obtained from the Fokker-Planck equation, respectively. The black and red circles represent the amplitudes of the first and third harmonics obtained from the Langevin function, respectively. The difference in harmonic amplitude between the Fokker-Planck equation and the Langevin function grows with increasing excitation frequency.

The simulation was conducted under different temperatures to estimate the temperature of MNPs using the temperature estimation model. In the simulation, the magnetic field strength was set to 0.001 T, and the excitation frequencies were set to 0.21 kHz, 1.1 kHz, and 2 kHz respectively. The temperature range was set from 310 K to 320 K with a step of 2 K. The MNPs were assumed to have a core diameter of 25 nm without the core size distribution, and the hydrodynamic size was 60 nm. The magnetization responses were respectively calculated from (1) and (4), and the amplitudes of the first (black squares and left axis) and third (red circles and right axis) harmonics were calculated using the DPSD algorithm under excitation frequencies of 0.21 kHz (Fig. 4(a)), 1.1 kHz (Fig. 4(b)), and 2 kHz (Fig. 4(c)). The harmonic amplitude of MNP magnetization decreases monotonically as the temperature increases.

The temperature was calculated from (9) and the amplitudes of the first and third harmonics under different temperatures using the least squares algorithm, and the true temperature was subtracted to give the temperature errors. Fig. 4(d) represents the temperature estimation errors under excitation frequencies of 0.21 kHz (black squares), 1.1 kHz (red circles), and 2 kHz (blue triangles) respectively. The maximum temperature error was less than 0.025 K in the temperature range from 310 K to 320 K. As depicted by the curves in Fig. 4(d), the temperature error increases as the excitation frequency increases. The temperature error at excitation frequencies of 2 kHz is larger than the temperature error at excitation frequencies of 0.21 kHz and 1.1 kHz, and the temperature error at an excitation frequency of 1.1 kHz is larger than the temperature error at an excitation frequency of 0.21 kHz.

As shown in Fig. 5, the compensation function was reconstructed using the estimated temperature under excitation frequencies of 0.21 kHz (Fig. 5(a)), 1.1 kHz (Fig. 5(b)), and 2 kHz (Fig. 5(c)). The black squares indicate the ratio of $Amp_{FP\text{-}1st}$ / $Amp_{FP\text{-}3rd}$ to $Amp_{L\text{-}1st}$ / $Amp_{L\text{-}3rd}$, $Amp_{FP\text{-}1st}$ / $Amp_{FP\text{-}3rd}$ represents the ratio of the first harmonic amplitude to the third harmonic amplitude obtained from the Fokker-Planck equation, $Amp_{L\text{-}1st}$ / $Amp_{L\text{-}3rd}$ represents the ratio of the first harmonic amplitude to the third harmonic amplitude obtained from the Langevin function, and the red circles represent the reconstructed compensation function ($G_1(a_1)/G_3(a_3)$).

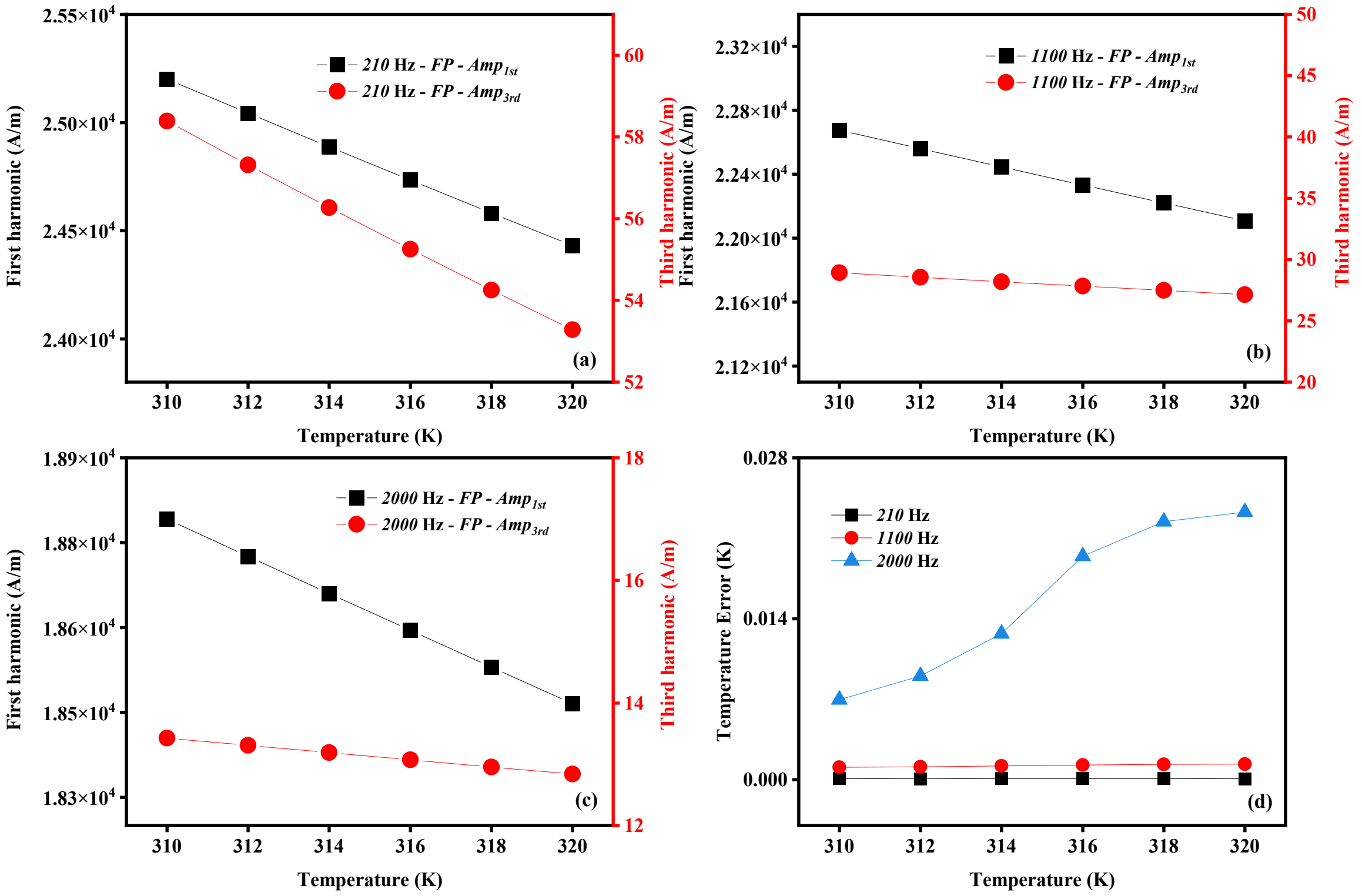

**Fig. 4.** Harmonic amplitudes were obtained from the Fokker-Planck equation under different temperatures. Amplitudes of the first (black squares) and third (red circles) harmonics were obtained under excitation frequencies of 0.21 kHz (a), 1.1 kHz (b), and 2 kHz (c) respectively. (d) Temperature error in the range of 310-320 K was obtained using (9) under excitation frequencies of 0.21 kHz (black squares), 1.1 kHz (red circles), and 2 kHz (blue triangles) respectively. $H_0$ = 0.001 T, $M_s$ = 200 kA/m, $d_c$ =25 nm, $d_h$ =60 nm, $\eta$ = 0.914× $10^{-3}$ Pas.

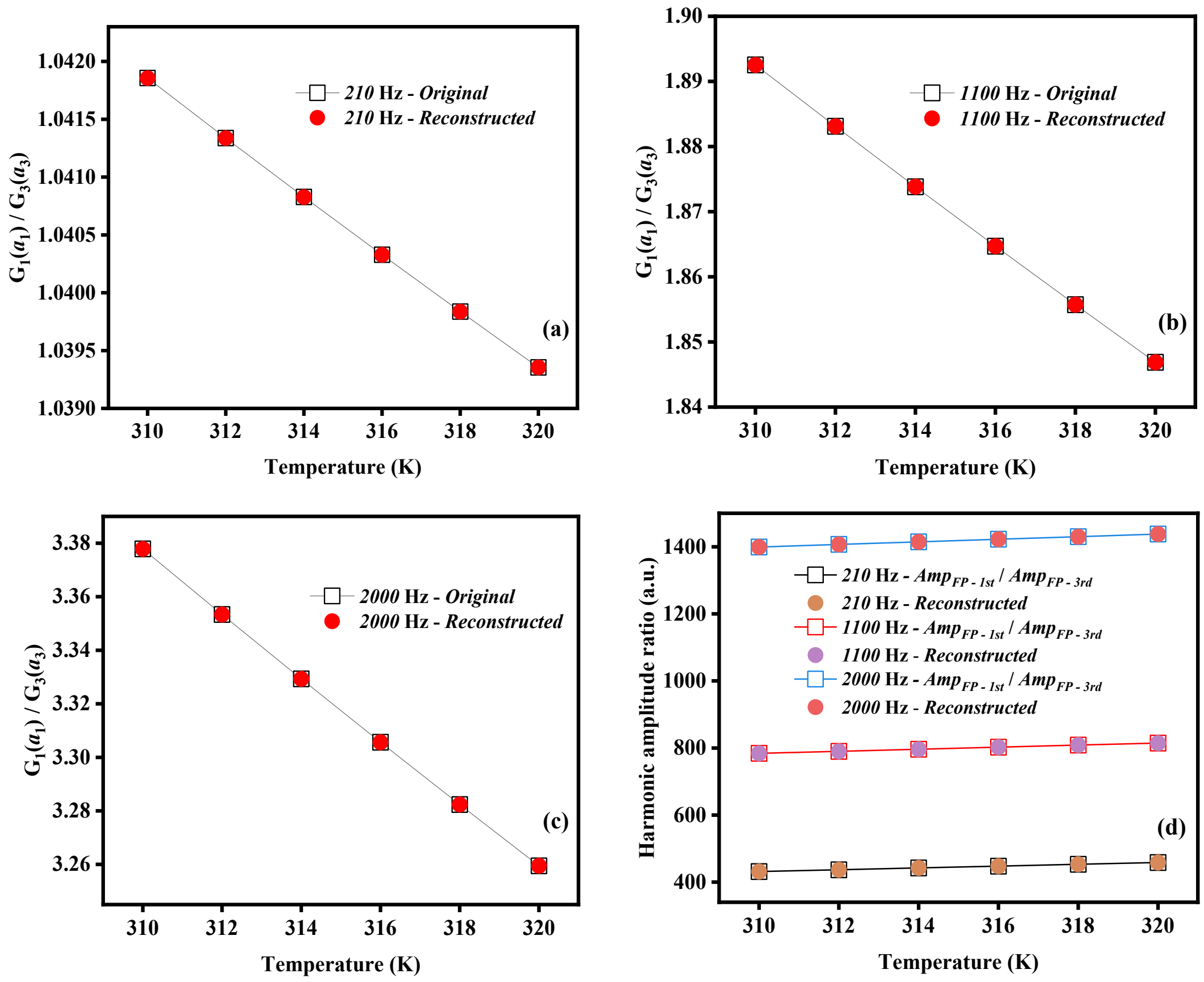

**Fig. 5.** The reconstructed compensation function under excitation frequencies of 0.21 kHz (a), 1.1 kHz (b), 2 kHz (c). (d) The reconstructed harmonic ratio was obtained utilizing (8) based on $G_1(a_1)/G_3(a_3)$ and the Langevin function. $H_0$ = 0.001 T, $d_c$ = 25 nm, $T$ was set from 310 K to 320 K with a step of 2 K, $M_s$ = 200 kA/m, $\eta = 0.914\times10^{-3}$ Pas.

Then we reconstructed the ratio of the amplitudes of the first to the third harmonics utilizing the reconstructed compensation function and the Langevin function under excitation frequencies of 0.21 kHz (brown circles in Fig. 5(d)), 1.1 kHz (purple circles in Fig. 5(d)), and 2 kHz (pink circles in Fig. 5(d)) respectively. As depicted by the curves in Fig. 5(d), the reconstructed ratio of the amplitudes of the first to the third harmonic agrees well with those of the harmonic ratio obtained from the Fokker-Planck equation.

## IV. Experiment And Results

### A. Experiment System

Fig. 6. shows the block diagram of the system with a magnetic nanothermometer based on magnetic particle spectroscopy (MPS), which is composed of an excited magnetic field generator and a weak magnetic signal measurement module. The excited magnetic field generator consists of a data acquisition card (DAQ) (NI-USB 6356, National Instruments, Austin, TX, USA), a power amplifier (AE 7224, AE Techron, Elkhart, IN, USA), and a pair of solenoid coils with the same structure and number of turns. The excitation magnetic field is generated by two solenoid coils in series (Exciting Coil 1 and Exciting Coil 2). The DAQ generates a sine signal, which is then inputted into the power amplifier for amplification, resulting in the solenoid coils to generate the alternating excitation magnetic field. The weak magnetic signal measurement module is composed of a differential coil that uses the same structure and size, and the differential coil is composed of a detection coil and a balance coil. The weak magnetic signal measurement module also includes a preamplifier, a DAQ, and the Labview software in the computer. Additionally, the residual magnetization produced by the excitation magnetic field will be coupled to the magnetization response of the MNPs. The residual magnetization will cause errors in the calculation of the first harmonic based on the Magnetization response of MNPs, and lead to temperature estimation errors of MNPs. The differential coil is employed to eliminate part of the residual magnetization, and the detection coil and the balance coil are connected with dotted terminals. The magnetization of MNPs under the excitation magnetic field is detected by the differential coil, and amplified by the preamplifier. Then the amplified magnetization response will be measured by DAQ and transmitted to the computer for processing. The magnetization harmonic of MNPs is calculated using the DPSD algorithm. Next, the MNP temperature is determined by (9) and the magnetization harmonic of MNPs.

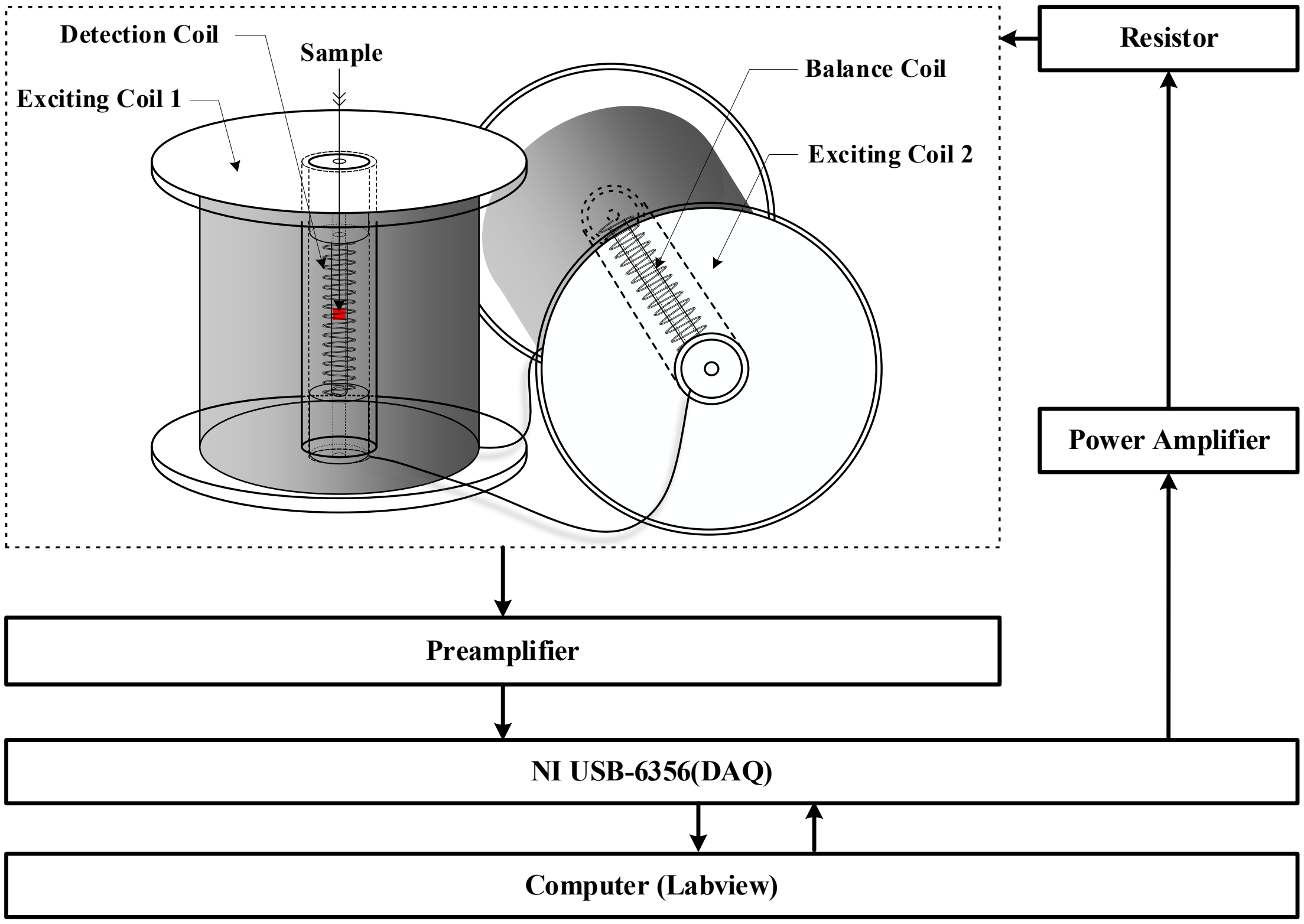

**Fig. 6.** The block diagram of the system with a magnetic nanothermometer based on MPS.

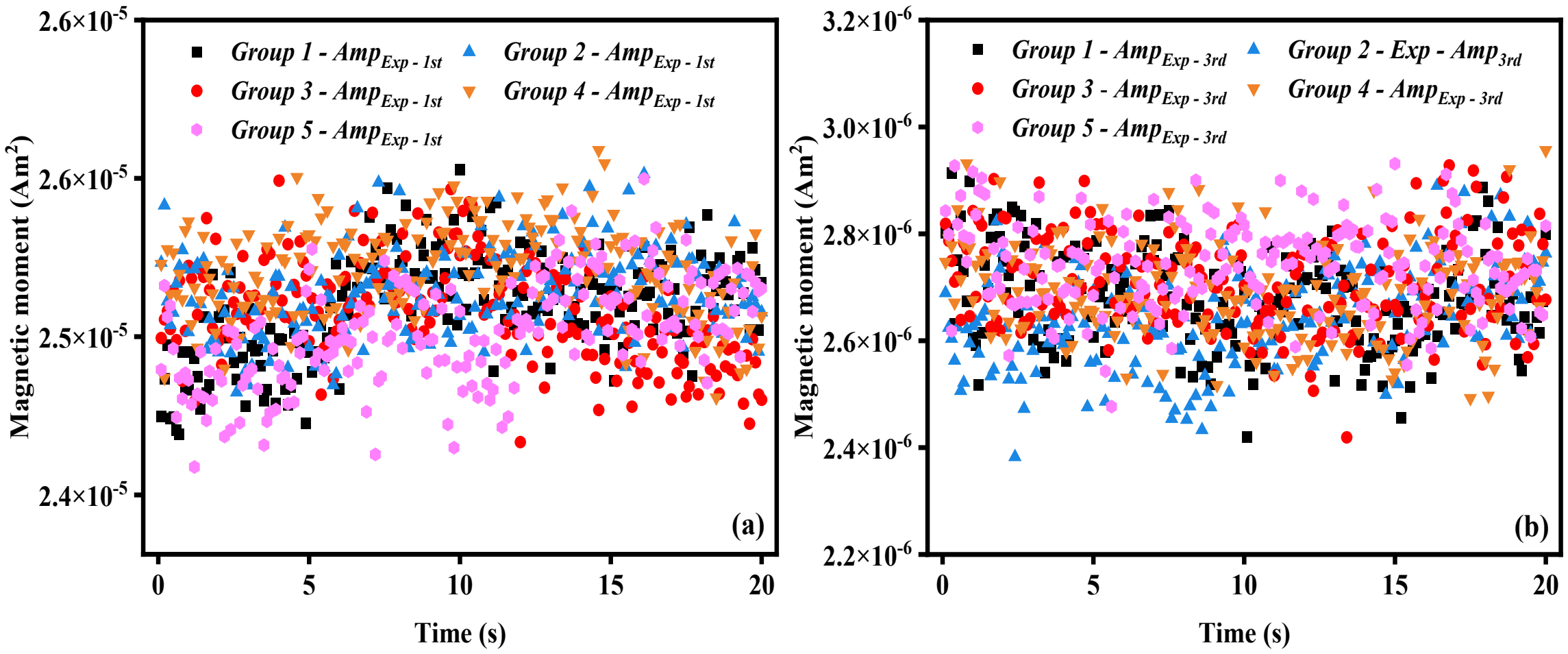

**Fig. 7.** (a) The amplitude of the first harmonic measured through five repeated experiments, and (b) the amplitude of the third harmonic measured through five repeated experiments. The excitation frequency was 0.21 kHz, and the magnetic field strength was 0.004 T.

We performed some experiments to examine the system, including the stability of the system, the lower detection limit of Fe, and the operating frequency range of the system. In the experiment, commercially available MNPs from Ocean NanoTech (SHP-30, San Diego, CA, USA) were used. These SHP-30 MNPs are specifically iron oxide nanoparticles with carboxylic acid groups, possessing an iron concentration of 5 mg/mL. The solvent of the sample is deionized $H_2O$ with 0.03% $NaN_3$. The experiments about the stability of the system were carried out with five repeated experiments. The excitation frequency was 0.21 kHz, and the magnetic field strength was 0.004 T. The MNP-based sample was put into the detection coil, and the amplitudes of the first and third harmonics of the MNP-based sample magnetization were measured in real-time using the system. We did five repeated experiments. Five repeated experiments of the first harmonic amplitude fluctuated in the range of $2.4204\times10^{-5}$ A m$^2$ and $2.57429\times10^{-5}$ A m$^2$ in Fig. 7(a), five repeated experiments of the third harmonic amplitude fluctuated in the range of $2.3827\times10^{-6}$ A m$^2$ and $2.931\times10^{-6}$ A m$^2$ in Fig. 7(b). The system satisfies the needs of the temperature measurement experiment.

We performed experiments to determine the lower detection limit of Fe, the excitation frequency was 0.21 kHz, and the magnetic field strength was 0.004 T. The amplitudes of the first (black squares and left axis), third (red squares and right axis (red)), and fifth (blue squares and right axis (blue)) harmonics were measured in real-time after the MNP-based sample had been placed in the detection coil. Eight experiments were performed using different weights of Fe (1 $\mu$g, 5 $\mu$g, 10 $\mu$g…), and the harmonic amplitudes are shown in Fig. 8(a). The amplitudes of harmonic increases with the weight of Fe. The detection range of the magnetization response signal of Fe of the system is from 1 $\mu$g to 150 $\mu$g. The lower detection limit of Fe is 1 $\mu$g, which meets the requirements for temperature measurement experiments.

To find the operating frequency range of the system, frequency sweep experiments were carried out under the excitation frequencies of 0.2 kHz, 0.51 kHz, 0.82 kHz, 1.21 kHz, 1.72 kHz, 2.22 kHz, 2.72 kHz, and 3.02 kHz, the magnetic field strength was 0.004 T. The amplitudes of the first (black squares and left axis), third (red squares and right axis (red)), and fifth (blue squares and right axis (blue)) harmonics were measured in real-time using the system. The amplitudes of the harmonic under different excitation frequencies are shown in Fig. 8(b). The amplitude of the harmonic diminishes as the excitation frequency increases. The higher the excitation frequency, the greater the effect of Brownian relaxation on harmonics. The operating frequency range of the system is 0.2 kHz to 3 kHz.

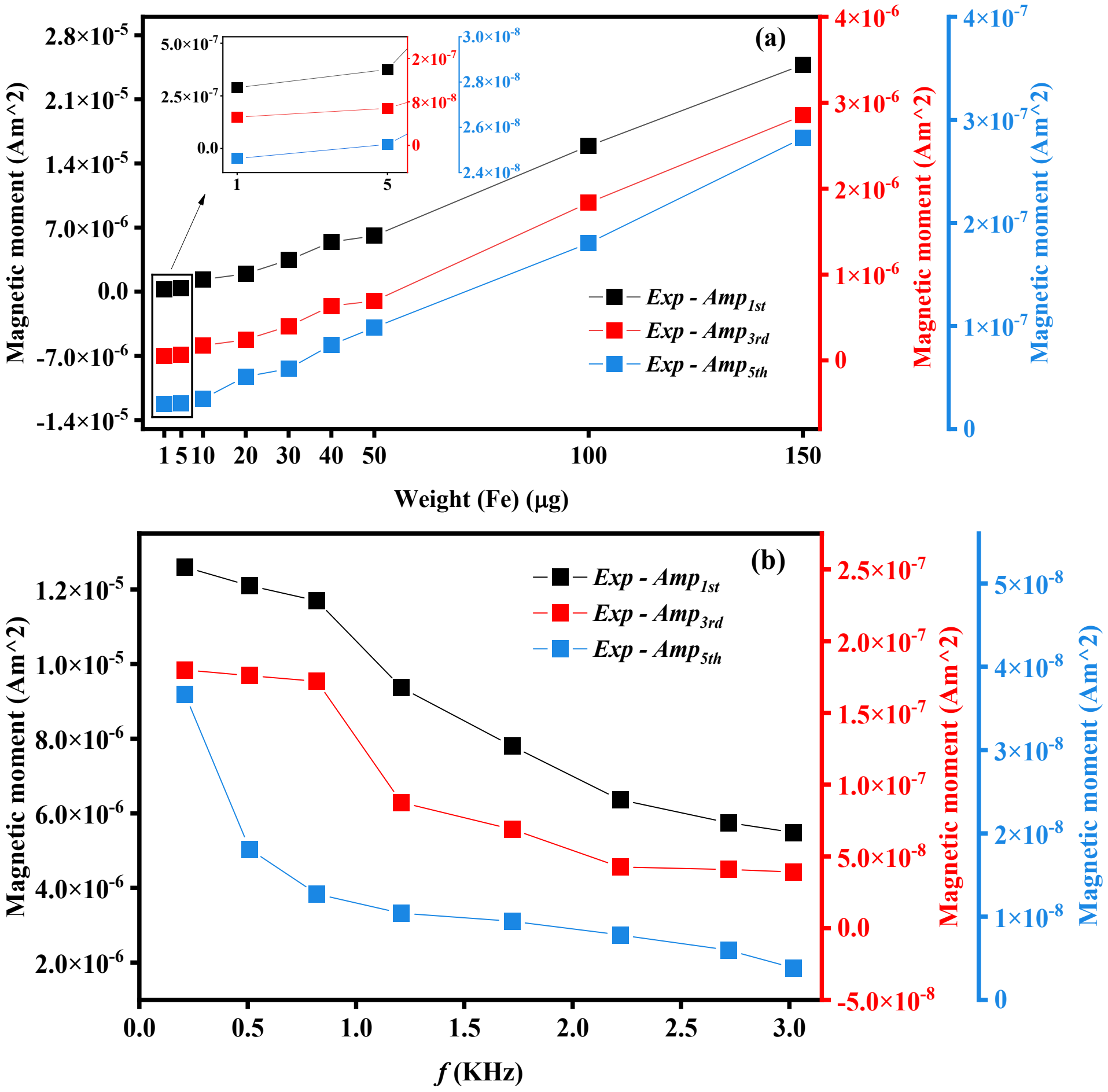

**Fig. 8.** (a) The lower detection limit of Fe of the system, and (b) the operating frequency range of the system. The black squares, the red squares, and the blue squares represent the amplitudes of the first (left axis), third (right axis (red)), and fifth (right axis (blue)) harmonics, respectively.

*B. Temperature Estimation*

Temperature experiments were carried out using the home-made system described above. In the experiments, the frequencies of AC magnetic fields are 0.21 kHz, 1.37 kHz, and 2.2 kHz respectively. The magnetic field strength was 0.0011 T, and the MNP-based sample used in the temperature experiments was SHP-30. A FOB100 optical fiber thermometer (produced by OMEGA) was inserted in the center of the MNP-based sample to monitor the temperature of the MNP-based sample, the MNP-based sample was heated to 322 K for water bath heating, and then put the MNP-based sample into the detection coil. During the natural cooling process of the MNP-based sample, the harmonic amplitudes of the MNP-based sample magnetization were measured in real-time at different temperatures using the system. The measured amplitudes of the first and third harmonics under AC magnetic fields in the temperature range of about 310-321 K are shown in Fig. 9.

Fig. 9 shows the harmonic amplitudes of the MNP magnetization under the excitation frequencies of 0.21 kHz

(squares in Fig. 9(a)), 1.37 kHz (squares in Fig. 9(b)), and 2.2 kHz (squares in Fig. 9(c)) respectively. The solid lines represent the fitting results of the measured harmonics. As depicted by the curves in Fig. 9(a), Fig. 9(b), and Fig. 9(c), the harmonic amplitude of the MNP magnetization decreases monotonically as the temperature increases. Using the fitting results and (9), as well as the reference temperature provided by the optical fiber thermometer, the temperature error can be assessed. Fig. 9(d) shows the error of temperature under excitation frequencies of 0.21 kHz (black squares), 1.37 kHz (red circles), and 2.2 kHz (blue triangles) respectively. The maximum error of temperature increases as the excitation frequency increases. The maximum error of temperature is about 0.007 K under an excitation frequency of 0.21 kHz, the maximum error of temperature increases to about 0.013 K under an excitation frequency of 1.37 kHz, and the maximum error of temperature increases to about 0.035 K under an excitation frequency of 2.2 kHz. The reason may be that the model only considers Brownian relaxation and ignores Néel relaxation. As the excitation frequency increases, the proportion of Brownian relaxation gradually decreases, while the proportion of Néel relaxation gradually increases. The model proposed in this study cannot accurately describe the magnetization harmonics of MNPs dominated by Néel relaxation, leading to temperature estimation errors.

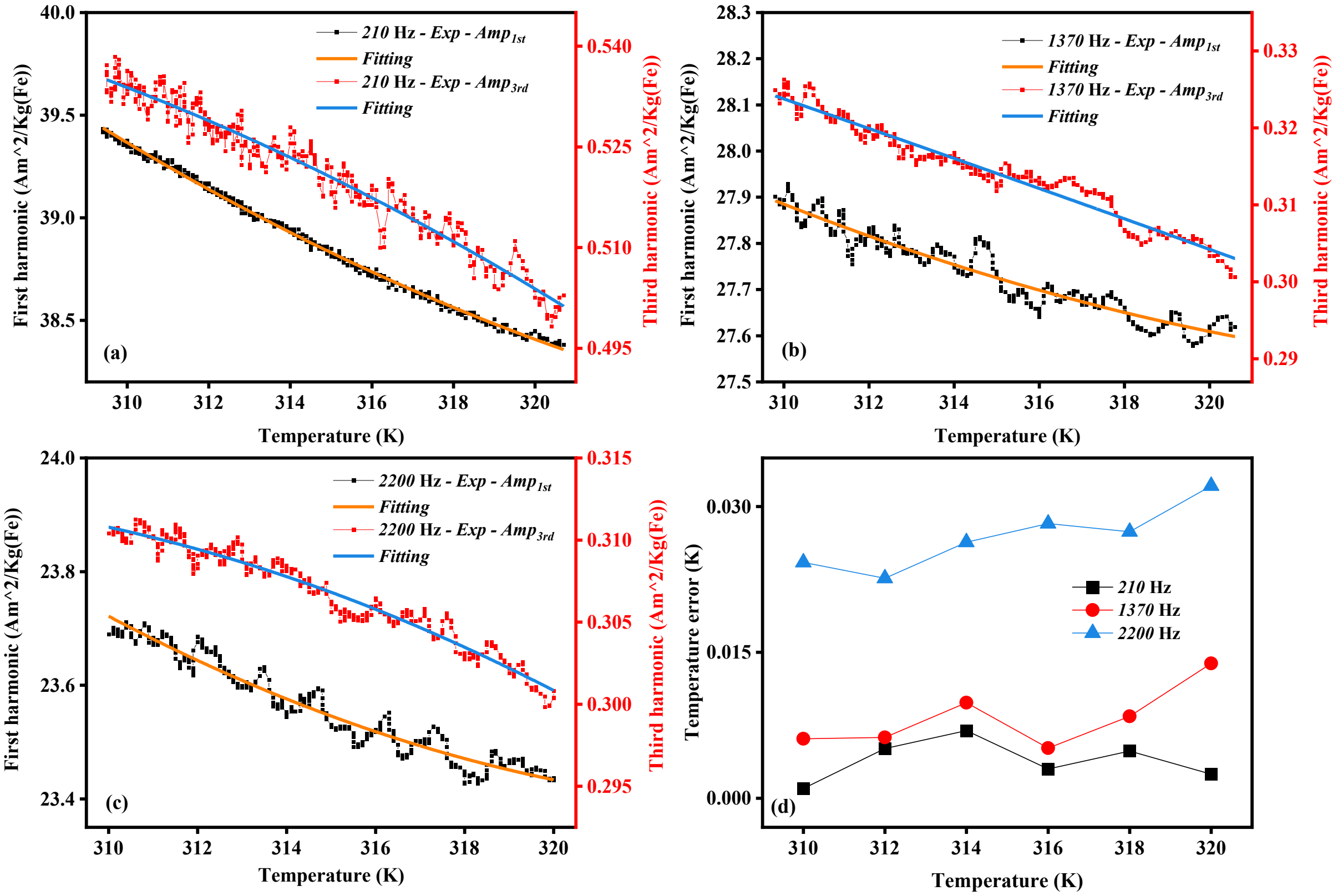

**Fig. 9.** Measured harmonic amplitudes (squares) and fitting results (solid lines) of measured harmonic amplitudes at different temperatures provided by an optical fiber thermometer. Amplitudes of the first and third harmonics were obtained under excitation frequencies of 0.21 kHz (a), 1.37 kHz (b), and 2.2 kHz (c) respectively. (d) Temperature error in the range of 310-320 K was obtained using (9) under excitation frequencies of 0.21 kHz (black squares), 1.37 kHz (red circles), and 2.2 kHz (blue triangles) respectively. The magnetic field strength was 0.0011 T.

As shown in Fig. 10. (a), (b), and (c), the compensation function was reconstructed using the estimated temperature under excitation frequencies of 0.21 kHz (Fig. 10. (a)), 1.37 kHz (Fig. 10. (b)), and 2.2 kHz (Fig. 10. (c)) respectively. The black squares indicate the ratio of $Amp_{Exp-1st}/Amp_{Exp-3rd}$ to $Amp_{L-1st}/Amp_{L-3rd}$, $Amp_{Exp-1st}/Amp_{Exp-3rd}$ represents the ratio of the first harmonic amplitude to the third harmonic amplitude measured in AC magnetic fields, $Amp_{L-1st}/Amp_{L-3rd}$ represents the ratio of the first harmonic amplitude to the third harmonic amplitude obtained from the Langevin function, and the red circles represent the reconstructed compensation function ($G_1(a_1)/G_3(a_3)$). Then we reconstructed the harmonic ratio utilizing the reconstructed compensation function and the Langevin function under excitation frequencies of 0.21 kHz (brown circles in Fig. 10. (d)), 1.37 kHz (purple circles in Fig. 10. (d)), and 2.2 kHz (pink circles in Fig. 10. (d)) respectively. As depicted by the curves in Fig. 10. (d), the reconstructed ratio of the amplitudes of the first to the third harmonic matches well with the experimental data.

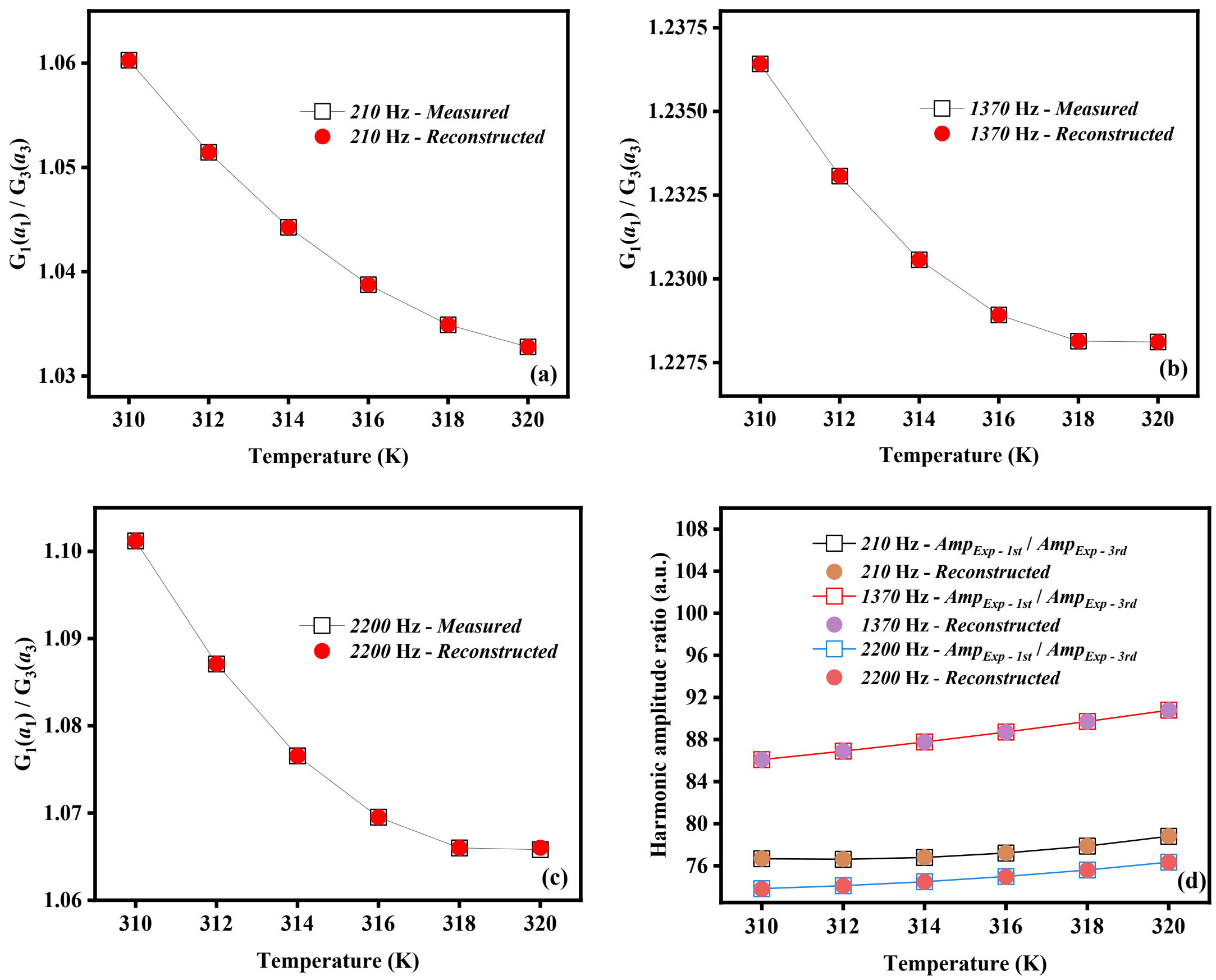

**Fig. 10.** The reconstructed compensation function under excitation frequencies of 0.21 kHz (a), 1.37 kHz (b), and 2.2 kHz (c) respectively. (d) The harmonic ratio was reconstructed utilizing (8) based on $G_1(a_1)/G_3(a_3)$ and the Langevin function.

## V. Discussion

This study presents a new method for estimating the temperature of MNPs based on the AC magnetization harmonic of MNPs dominated by Brownian relaxation. Some factors affect the temperature measurement precision of the system, which can be improved mainly from two aspects.

First, the output signal of the detection coil is induced not only by the MNP magnetization but also by the applied AC magnetic field, named residual magnetic field. The signal produced by the residual magnetic field is named the residual magnetization. To suppress the residual magnetization induced by the residual magnetic field, the detection coil and the balance coil are connected with dotted terminals in Fig. 6. It is not easy to clear away completely the residual magnetization generated by the residual magnetic field. The measurement precision of the system can be improved by suppressing the residual magnetic field down to a noise level.

Second, the residual magnetization generated by the residual magnetic field will cause errors in the calculation of the first harmonic based on the Magnetization response of MNPs, and clearing away residual magnetization completely is no easy task. High-order harmonics are measured in the AC magnetic field for MNPT instead of the first harmonic. The high-order harmonics decay more rapidly than the low-order ones. There are high requirements for the signal sensitivity of the system. The signal strength of the high-order harmonic increases as the magnetic field strength increases. The exciting coils would heat up due to excessive magnetic field strength, which will cause great interference during the measurement process. To overcome this excessive heating in the coils, it is necessary to add a cooling device to the coils.

## VI. Conclusion

In this paper, we report on a novel estimation method for the temperature of MNPs dominated by Brownian relaxation. To establish a simple temperature estimation model, we studied the difference in the AC magnetization response and magnetization harmonics between the Fokker-Planck equation and the Langevin function, and analyzed the relationship between the harmonic amplitude and the key factors, the key factors include Brownian relaxation time, temperature, magnetic field strength, core size and hydrodynamic size of MNPs, excitation frequency, and so on. According to the difference between the Fokker-Planck equation and the Langevin function, we proposed a compensation function for AC magnetization harmonic. The compensation function also took into account the key factors. Then a temperature estimation model based on the compensation function and the Langevin function was established, and the temperature estimation model was solved using the least squares algorithm. The experimental results allow for an accuracy of the temperature probing of 0.035 K in the temperature range from 310 K to 320 K. The temperature estimation model contributes to the improvement of the performance of MNPT, and is expected to be applied to MNPH.